\documentclass[reprint,amsmath,amssymb,aps,prl,superscriptaddress]{revtex4-2}

\usepackage{graphicx}
\usepackage{bm}
\usepackage[dvipsnames]{xcolor}
\usepackage[
  colorlinks=true,
  bookmarks=false,
  citecolor=NavyBlue,
  linkcolor=NavyBlue,
  urlcolor=NavyBlue,
  pdftitle={Sign-problem-resilient singular-value probe in determinant quantum Monte Carlo},
  pdfauthor={Wen Chen and Rubem Mondaini}
]{hyperref}
\usepackage{orcidlink}
\usepackage{mathrsfs}
\graphicspath{{../figures/}}

\newcommand{\prlsection}[1]{\noindent\textit{#1.---}\hspace{0.25em}}

\newcommand{\rmi}{\mathrm{i}}
\newcommand{\beginsupplement}{%
  \setcounter{section}{0}%
  \renewcommand{\thesection}{S\arabic{section}}%
  \setcounter{equation}{0}%
  \renewcommand{\theequation}{S\arabic{equation}}%
  \setcounter{table}{0}%
  \renewcommand{\thetable}{S\arabic{table}}%
  \setcounter{figure}{0}%
  \renewcommand{\thefigure}{S\arabic{figure}}%
}

\begin{document}

\title{Sign-problem-resilient singular-value probe in determinant quantum Monte Carlo}

\author{Wen Chen\,\orcidlink{0000-0001-6089-9909}}
\affiliation{Department of Physics, University of Houston, Houston, Texas 77004, USA}

\author{Rubem Mondaini\,\orcidlink{0000-0001-8005-2297}}
\email{rmondaini@uh.edu}
\affiliation{Department of Physics, University of Houston, Houston, Texas 77004, USA}
\affiliation{Texas Center for Superconductivity, University of Houston, Houston, Texas 77004, USA}


\begin{abstract}
The sign problem limits determinant quantum Monte Carlo studies of strongly correlated fermion systems. In the spin-channel Hubbard-Stratonovich decoupling, spin correlations are exactly related to auxiliary-field correlations. This relation implies that an antiferromagnetic transition reorganizes auxiliary-field configurations and thereby changes the statistical structure of the resulting fermion matrices. We use the adjacent gap ratio of low-lying singular values of the space-time fermion matrix to probe interaction-driven transitions in two half-filled honeycomb-lattice Hubbard models. In the sign-free honeycomb Hubbard model, the statistic tracks the established transition from a Dirac semimetal to an antiferromagnetic Mott insulator. In the complex-weight Haldane-Hubbard model, the transition-sensitive feature remains visible in the phase-quenched reference ensemble and occurs near previous estimates of the transition. Moreover, phase reweighting only weakly modifies the gap ratio over the regimes investigated, despite the rapid suppression of the average phase. These results establish singular-value statistics as a sign-problem-resilient probe of interaction-driven transitions in determinant quantum Monte Carlo.
\end{abstract}

\maketitle


\prlsection{Introduction} Determinant quantum Monte Carlo (DQMC) methods have been instrumental in unveiling the physics of various fermionic many-body models, encompassing conventional symmetry-breaking phases~\cite{Hirsch1985, Meng2010, sorella2012absence, assaad2013pinning} as well as interaction effects in topological systems~\cite{hohenadler2011correlation, hohenadler2012quantum, zheng2011particle, Wang2014}. It does so via a Hubbard-Stratonovich (HS) transformation~\cite{hirsch1983discrete}, which maps an interacting problem onto quadratic fermions coupled to auxiliary fields, providing an unbiased numerical method for strongly correlated systems~\cite{blankenbecler1981monte,scalapino1981monteII}. The ensuing weights, written as determinants of fermion matrices, can nonetheless be negative or complex. In this case, the Monte Carlo sampling instead uses their magnitudes to define a reference ensemble, while physical averages require sign or phase reweighting~\cite{Hirsch1985}. Yet, the average reweighting factor typically decreases exponentially with inverse temperature and system size~\cite{loh1990sign,iglovikov2015geometry}, causing the computational effort required to maintain a fixed relative precision to grow exponentially~\cite{Troyer2005}.

Nevertheless, although reweighting becomes exponentially costly, configurations in the reference ensemble can still be generated with the usual polynomial matrix-algebra cost per DQMC update~\cite{Huang2018,Huang2023,Zhao2025}. Indeed, transition information has been extracted from this ensemble using convolutional neural networks trained on equal-time Green's-function matrices~\cite{broecker2017machine}, Hamming distances between auxiliary-field configurations~\cite{yi2022hamming}, and the average sign or phase factor itself~\cite{mondaini2022quantum,mondaini2023universality,ma2024universal}. More recent work has further shown that the bias introduced by omitting reweighting depends strongly on the observable under consideration~\cite{larson2026signresolved}. These results raise a natural question: can the matrix structure generated within DQMC provide a simple and interpretable diagnostic of phase transitions that is only weakly affected by sign or phase reweighting?

Here, we address this question through the singular-value statistics of the configuration-dependent fermion matrices that emerge on the sampling. The singular values are directly tied to the reference measure: their product determines the magnitude of the corresponding determinant, whereas their local spectral correlations~\cite{mehta2004random,haake2010quantum} contain information not retained by this product alone. We show that these correlations, quantified through the ratio of consecutive singular-value gaps~\cite{atas2013distribution,kawabata2019symmetry,kawabata2023singular}, track the underlying phase transitions. Remarkably, the resulting average gap ratio is only weakly modified by sign or phase reweighting over the regimes investigated. Thus, although the singular values do not encode the sign or phase of an individual determinant, their local correlations provide a transition diagnostic that is largely shared by the reference and reweighted measures.

We first benchmark this approach in the sign-problem-free half-filled Hubbard model on the honeycomb lattice, which undergoes a transition from a Dirac semimetal to an antiferromagnetic (AFM) Mott insulator as the interaction strength increases~\cite{sorella2012absence,assaad2013pinning,parisentoldin2015fermionic,Buividovich2019, Ostmeyer2020, Ostmeyer2021, mondaini2022quantum,mondaini2023universality}. We then turn to the half-filled Haldane-Hubbard model, whose complex determinant weights give rise to a phase problem~\cite{imriska2016first,vanhala2016topological,he2024phase}. Even when evaluated entirely within the phase-quenched reference ensemble, the gap-ratio feature occurs at an interaction strength consistent with an exact-diagonalization benchmark on the same finite cluster and close to previous many-body estimates of the transition~\cite{imriska2016first,vanhala2016topological,he2024phase}.

\begin{figure}[t]
  \centering
  \includegraphics[width=\columnwidth]{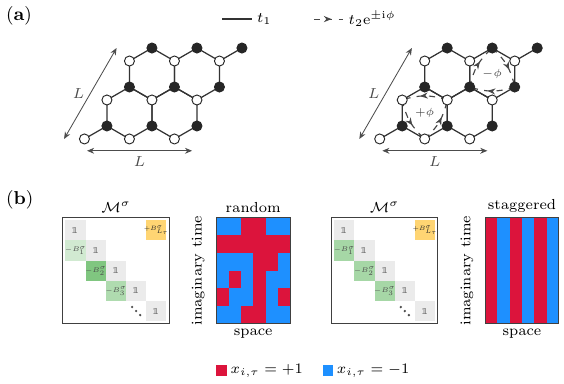}
  \caption{Schematic of the models and DQMC space-time fermion matrix $\mathcal M^\sigma$. (a) Honeycomb Hubbard model with $t_2=0$ (left) and Haldane-Hubbard model with complex next-nearest-neighbor hopping $t_2e^{\rmi\phi_{ij}}$ (right). Open and filled circles denote the two sublattices, and the arrows specify the phase convention. (b) Sparse block structure of $\mathcal M^\sigma$ and auxiliary fields on the $N_s\times L_\tau$ space-time lattice. The two fields illustrate a random weak-coupling background and an idealized staggered, $\tau$-uniform strong-coupling limit, respectively.}
  \label{fig:models-matrix}
\end{figure}

\prlsection{Models, fermion matrices, and field structure} We consider spin-$\tfrac12$ fermions on the honeycomb lattice at half-filling, described by the Hamiltonian
\begin{align}
  \hat H={}&-t_1\sum_{\langle ij\rangle,\sigma}
  \hat c^\dagger_{i\sigma}\hat c_{j\sigma}^{\phantom{\dagger}}
  -t_2\sum_{\langle\!\langle ij\rangle\!\rangle,\sigma}
  e^{\rmi\phi_{ij}}\hat c^\dagger_{i\sigma}\hat c_{j\sigma}^{\phantom{\dagger}}
  \nonumber\\
  &+U\sum_i
  \left(\hat n_{i\uparrow}-\tfrac12\right)
  \left(\hat n_{i\downarrow}-\tfrac12\right).
  \label{eq:hamiltonian}
\end{align}
Here, $\hat c^\dagger_{i\sigma}$ ($\hat c_{i\sigma}^{\phantom{\dagger}}$) creates (annihilates) a spin-$\sigma$ fermion at site $i$, $\hat n_{i\sigma}=\hat c^\dagger_{i\sigma}\hat c_{i\sigma}^{\phantom{\dagger}}$, and $U>0$ is the onsite repulsion strength. The sums over $\langle ij\rangle$ and $\langle\!\langle ij\rangle\!\rangle$ run over nearest- and next-nearest-neighbor (NNN) bonds  with hopping amplitudes $t_1$ and $t_2e^{\rmi\phi_{ij}}$, respectively, where the Peierls phase satisfies $\phi_{ji}=-\phi_{ij}$. We take $t_1$ as the energy unit and focus on the two cases shown in Fig.~\ref{fig:models-matrix}(a): the honeycomb Hubbard model with $t_2=0$, and the Haldane-Hubbard model with $t_2/t_1=0.2$ and $\phi_{ij}=\pm\pi/2$. We use periodic lattices with $L\times L$ unit cells and $N_s=2L^2$ sites.

In finite-temperature DQMC~\cite{santos2003introduction,assaad2008worldline,bai2009numerical,gubernatis2016quantum}, we discretize the inverse temperature as $\beta=L_\tau\Delta\tau$ and introduce binary auxiliary fields $x_{i,\tau}=\pm1$ through a discrete spin-channel HS transformation~\cite{hirsch1983discrete}. After integrating out the fermions, each auxiliary-field configuration $x$ has weight $W[x]=\prod_\sigma\det M^\sigma[x]$, where $M^\sigma[x]=I+B_{L_\tau}^\sigma[x]\cdots B_1^\sigma[x]$. The single-slice propagator is $B_\tau^\sigma[x]=e^{\sigma\lambda V_\tau[x]}e^{-\Delta\tau K}$. Here, $K$ is the single-particle kinetic matrix corresponding to the hopping terms, $V_\tau[x]=\operatorname{diag}(x_{1,\tau},\ldots,x_{N_s,\tau})$, and the HS coupling $\lambda$ satisfies $\cosh\lambda=e^{\Delta\tau U/2}$. At half filling, the honeycomb Hubbard model is sign-problem-free, whereas the complex NNN hopping in the Haldane-Hubbard model gives rise to generally complex weights, $W[x]=|W[x]|e^{\rmi\theta[x]}$, wherein we sample auxiliary-field configurations using $W_{\rm ref}[x]=|W[x]|$.

To avoid explicitly forming the ill-conditioned long product in $M^\sigma$ at low temperatures~\cite{loh1989stable,loh2005numerical,bauer2020fast}, we analyze the $N_sL_\tau\times N_sL_\tau$ sparse space-time fermion matrix $\mathcal M^\sigma[x]$ shown in Fig.~\ref{fig:models-matrix}(b). It possesses only $2L_\tau$-nonzero $N_s\times N_s$ blocks, and satisfies $\det\mathcal M^\sigma[x]=\det M^\sigma[x]$~\cite{blankenbecler1981monte}. Sampling, stabilization, and numerical details are given in the Supplemental Material (SM)~\cite{supplementalMaterial}.

Notably, at fixed $\Delta\tau$, $\lambda$ grows monotonically with $U$, thereby increasing the magnitude of the auxiliary-field term $\sigma\lambda V_\tau[x]$ in $B_\tau^\sigma[x]$. In turn, at distinct space-time points, the auxiliary-field and $\hat m_i\equiv\hat n_{i\uparrow}-\hat n_{i\downarrow}$ correlations, both evaluated with the physical weights, obey~\cite{hirsch1983discrete}
\begin{equation}
  \bigl\langle x_{i,\tau}x_{j,\tau'}\bigr\rangle_{\rm phys}
  =
  \tanh^2\lambda\,
  \bigl\langle
    \hat m_i(\tau)\hat m_j(\tau')
  \bigr\rangle_{\rm phys}.
  \label{eq:hs-spin-correlation}
\end{equation}
Here $\tanh^2\lambda$ rescales only the overall magnitude, while the spatial and imaginary-time structure of the auxiliary-field correlations is inherited from the physical spin correlations.

In the weak-coupling regime ($\lambda \ll 1$), the system has not yet developed AFM order. The fields therefore form a nearly uncorrelated binary background over space and imaginary time, as illustrated on the left of Fig.~\ref{fig:models-matrix}(b), and $e^{\sigma\lambda V_\tau}$ remains close to the identity. On the other hand, as $U$ increases into the AFM Mott regime, the spin correlations develop a staggered spatial structure and persist over increasingly long imaginary-time intervals, while the auxiliary-field correlations evolve, at the ensemble level, toward the idealized limit shown on the right of Fig.~\ref{fig:models-matrix}(b). The interactions therefore change both the magnitude of the auxiliary-field term and the spatiotemporal organization of the field. Since both effects enter directly into the space-time matrix $\mathcal M^\sigma$, we next ask how they are reflected in its low-lying singular-value correlations.

\begin{figure}[t]
  \centering
  \includegraphics[width=\columnwidth]{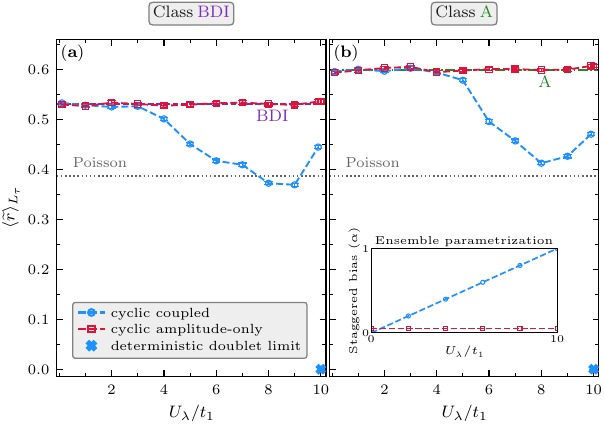}
\caption{Low-lying singular-value statistic $\langle\widetilde r\rangle_{L_\tau}$ for the synthetic probability-flip ensembles in (a) class BDI and (b) class A, with $L=6$ and $L_\tau=80$. At each space-time point, the binary field equals the staggered template with probability $(1+\alpha)/2$ and its opposite with probability $(1-\alpha)/2$. Along the amplitude-only path, $\alpha=0.05$ is fixed while $U_\lambda$ changes only the coupling strength $\lambda(U_\lambda)$. Along the coupled path, $\alpha=U_\lambda/U_\star$, with $U_\star/t_1=10$, so the field becomes progressively biased toward the staggered template (inset). Dashed (dash-dotted) horizontal lines denote the BDI (class-A) random-matrix benchmark, and the dotted line marks the Poisson value. At the deterministic endpoint $U_\lambda=U_\star$ ($\alpha=1$), pairwise singular-value doublets drive $\langle\widetilde r\rangle_{L_\tau}$ to zero; this point is shown separately from the finite-flip coupled trajectory. Error bars denote the standard error over independent realizations.
}
  \label{fig:cyclic-ensemble}
\end{figure}

\prlsection{Singular-value statistics in synthetic matrix ensembles} To disentangle these two effects, we construct synthetic ensembles with the same block-cyclic structure as $\mathcal M^\sigma$, in which the auxiliary-field coupling strength and its staggered bias can be controlled independently. For each realization, a single kinetic matrix $K$ is drawn and kept fixed along imaginary time, using either $K_{\rm BDI}=\left(\begin{smallmatrix}0&T\\ T^{\mathsf T}&0\end{smallmatrix}\right)$ with $T$ a real Gaussian random matrix~\cite{Mehta2004}, or a generic complex Hermitian Gaussian matrix $K_{\rm A}$, representing the two symmetry classes relevant to the honeycomb Hubbard and Haldane-Hubbard kinetic terms, respectively~\cite{kawabata2019symmetry,kawabata2023singular}.

The binary auxiliary field is generated relative to the staggered template $\eta_i=\pm1$ on the two sublattices. At every space-time point independently, we choose $x_{i,\tau}=\eta_i$ with probability $(1+\alpha)/2$ and $x_{i,\tau}=-\eta_i$ with probability $(1-\alpha)/2$, so that $\alpha$ controls the staggered bias without introducing connected space-time correlations. The synthetic parameter $U_\lambda$ controls the field-coupling strength through $\cosh\lambda=\exp(\Delta\tau U_\lambda/2)$, following the same parametrization as the Hubbard-model HS transformation.

Hereafter, we restrict the spectral analysis to the spin-up sector and suppress the spin label. For each matrix realization, we order the singular values of $\mathcal M^\uparrow$ as $0\le s_1\le s_2\le\cdots$, define $\delta_n=s_{n+1}-s_n$, and use the adjacent gap ratio~\cite{Oganesyan2007,atas2013distribution,kawabata2023singular}
\begin{equation}
  \widetilde r_n
  =
  \frac{\min\!\left(\delta_n,\delta_{n+1}\right)}
       {\max\!\left(\delta_n,\delta_{n+1}\right)}\ ,
  \label{eq:adjacent-gap-ratio}
\end{equation}
as a metric of local spectral correlations. We report $\langle\widetilde r\rangle_{L_\tau}$, obtained by averaging $\widetilde r_n$ over the lowest $L_\tau$ singular values and then over the matrix ensemble; its dependence on the window size is examined in the SM~\cite{supplementalMaterial}.

Figure~\ref{fig:cyclic-ensemble} compares two trajectories through the synthetic parameter space $(U_\lambda,\alpha)$. Along the amplitude-only path, $\alpha=0.05$ is fixed, so the field distribution is unchanged while only the coupling strength $\lambda(U_\lambda)$ increases. In both symmetry classes, $\langle\widetilde r\rangle_{L_\tau}$ remains close to the corresponding random-matrix benchmark over the entire path. Along the coupled path, instead, $\alpha=U_\lambda/U_\star$ with $U_\star/t_1=10$, so increasing $U_\lambda$ simultaneously strengthens the field coupling and biases the binary configurations toward the staggered template. The gap ratio then departs strongly from its random-matrix value in both classes. Since the two paths have the same $\lambda$ at each $U_\lambda$ and differ only in the auxiliary-field distribution, their separation shows that increasing the coupling amplitude alone is insufficient to generate the low-lying spectral response; a reorganization of the field distribution is essential.

\begin{figure}[t]
  \centering
  \includegraphics[width=\columnwidth]{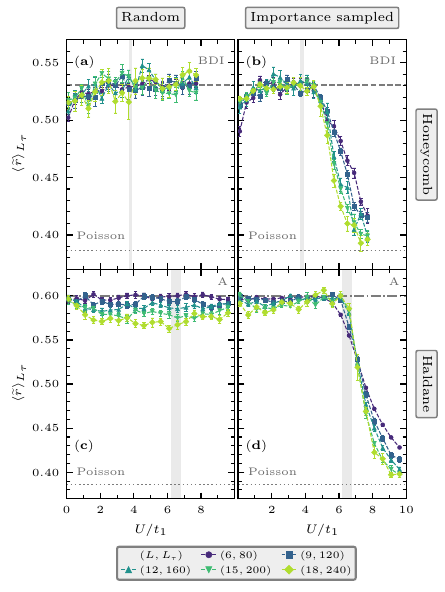}
  \caption{Mean adjacent-gap ratio $\langle\widetilde r\rangle_{L_\tau}$ of the low-lying singular values for the two lattice models. The top and bottom rows show the honeycomb Hubbard and Haldane-Hubbard models, respectively. The left column retains each model's $K$ and $\lambda(U)$ but uses independent, equiprobable fields $x_{i,\tau}=\pm1$ at every space-time point, whereas the right column uses fields generated by DQMC importance sampling. Curves are labeled by $(L,L_\tau)$. Gray bands mark the literature estimates $U_c/t_1\in[3.7,3.9]$ for the honeycomb Hubbard model  \cite{sorella2012absence,assaad2013pinning,parisentoldin2015fermionic} and $U_c/t_1\in[6.2,6.8]$ for the Haldane-Hubbard model~\cite{vanhala2016topological,he2024phase}. Horizontal lines indicate the BDI and class-A random-matrix benchmarks for the top and bottom rows, respectively~\cite{kawabata2023singular}, together with the Poisson value~\cite{atas2013distribution}. Error bars denote one standard error over independent randomized-field realizations in (a,c) and independent Markov-chain means in (b,d).}
  \label{fig:benchmark}
\end{figure}


\prlsection{Singular-value statistics in DQMC}Having established in controlled ensembles that singular-value correlations respond to the organization of the auxiliary field, we now test this response in DQMC. We compare fields generated by importance sampling with a control ensemble in which $x_{i,\tau}=\pm1$ is chosen randomly at every space-time point, while keeping the same kinetic matrix $K$ and HS coupling $\lambda(U)$.  For the size scan in Fig.~\ref{fig:benchmark}, we take $L_\tau\propto L$ at fixed $\Delta\tau$, thereby assuming a $z=1$ dynamic critical exponent, as expected for the Gross-Neveu critical point of model~\cite{parisentoldin2015fermionic}.

In Fig.~\ref{fig:benchmark}(a), the randomized-field results rapidly approach the corresponding random-matrix benchmark as $U$ increases and remain nearly unchanged thereafter. By contrast, the DQMC results in Fig.~\ref{fig:benchmark}(b) gradually depart from the benchmark after crossing the shaded critical interval~\cite{Meng2010, sorella2012absence, assaad2013pinning,parisentoldin2015fermionic} and decrease on the strong-coupling side, thereby tracking the established transition from a Dirac semimetal to an AFM Mott insulator. Because the two ensembles differ only in the joint distribution of the auxiliary field, their separation shows that this spectral response cannot be explained by $K$ and $\lambda(U)$ alone but depends on the field organization generated by the fermion-determinant weight.

We repeat this comparison for the complex-weight Haldane-Hubbard model, using the same $(L,L_\tau)$ sequence for direct comparison and sampling the phase-quenched measure $W_{\rm ref}=|W|$. As in the honeycomb case, the independently randomized fields remain close to the corresponding random-matrix value [Fig.~\ref{fig:benchmark}(c); window-size dependence is analyzed in the SM~\cite{supplementalMaterial}]. The reference-ensemble results, however, undergo a pronounced downward crossover whose steepest region overlaps the previously reported interaction range for the Chern insulator-to-Mott insulator transition [Fig.~\ref{fig:benchmark}(d)], which comes from calculations in cylinders using density matrix renormalization group methods~\cite{he2024phase} or small-cluster dynamical mean-field theory results~\cite{vanhala2016topological}. Thus, a transition-sensitive reorganization of the low-lying singular-value correlations remains visible entirely within the phase-quenched reference ensemble, without phase reweighting.

These results establish that the phase-quenched matrix ensemble retains a transition-sensitive singular-value signal. At the same time, determination of whether its ensemble average coincides with the physically reweighted value is still an open question. We address this next by expressing the reweighting bias in terms of the covariance between $\widetilde r$ and the determinant phase.

\prlsection{Phase-reweighting bias of the gap ratio}
To clarify the relation between the singular-value statistic and the phase-quenched measure, consider the polar decomposition of the space-time fermion matrix for each auxiliary-field configuration,
\begin{equation}
  \mathcal M^\sigma[x]
  =
  Q^\sigma[x]R^\sigma[x],
  \qquad
  R^\sigma[x]
  =
  \bigl(\mathcal M^{\sigma\dagger}[x]
  \mathcal M^\sigma[x]\bigr)^{1/2}.
\end{equation}
The determinant of $Q^\sigma$ has unit modulus and carries the phase, whereas the eigenvalues of the positive-semidefinite matrix $R^\sigma$ are the singular values $s_n^\sigma$. Consequently, $W[x] = \prod_\sigma\det\mathcal M^\sigma[x] = e^{\rmi\theta[x]} \prod_{\sigma,n}s_n^\sigma[x]$. Thus, the reference weight $|W[x]|=\prod_{\sigma,n}s_n^\sigma[x]$ is determined by the product of the singular values, whereas $\widetilde r_{L_\tau}[x]$ probes their local spacing correlations, often used to classify non-Hermitian chaotic matrices~\cite{kawabata2023singular}. Indeed, the determinant phase does not explicitly enter the gap ratio of a fixed configuration. Nonetheless, this configuration-level separation does not imply statistical independence between $\widetilde r_{L_\tau}[x]$ and $e^{\rmi\theta[x]}$, since both are determined by the same auxiliary-field configuration.
\begin{figure}[t]
  \centering
  \includegraphics[width=\columnwidth]{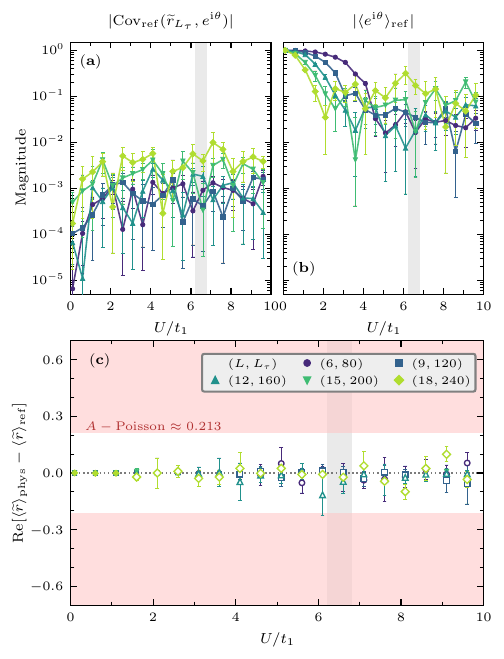}
  \caption{Phase-reweighting diagnostics for the Haldane-Hubbard model. (a) Magnitude of the reference-ensemble covariance between the low-lying singular-value statistic and the determinant phase, $|\mathrm{Cov}_{\rm ref}(\widetilde r_{L_\tau},e^{\rmi\theta})|$. (b) Magnitude of the average phase, $|\langle e^{\rmi\theta}\rangle_{\rm
  ref}|$. (c) Reweighting correction, $\mathrm{Re}[\langle\widetilde r\rangle_{L_\tau,\mathrm{phys}}-\langle\widetilde r\rangle_{L_\tau,\mathrm{ref}}]$. Error bars in (a,b) denote one standard error; those in (c) are obtained from a synchronized chain-block bootstrap. Filled symbols in (c) denote statistically resolved average-phase denominators, whereas open symbols are unresolved and shown only as diagnostics. The gray band marks the transition interval $U/t_1\in[6.2,6.8]$~\cite{vanhala2016topological,he2024phase}. Red regions in (c) mark corrections whose magnitude exceeds the separation between the class-A and Poisson benchmarks, $\langle\widetilde r\rangle_{\rm A}-\langle\widetilde r\rangle_{\rm P}\simeq0.213$~\cite{kawabata2023singular,atas2013distribution}.}

  \label{fig:covariance}
\end{figure}

Whether phase reweighting changes the ensemble average of $\widetilde r_{L_\tau}$ is therefore controlled not by the polar decomposition itself, but by the statistical correlation between the gap ratio and the determinant phase. Generalizing the sign-reweighting identity of Ref.~\cite{larson2026signresolved} to complex weights, the physical reweighted and reference-ensemble averages satisfy
\begin{equation}
  \langle \widetilde r_{L_\tau}\rangle_{\rm phys}
  -
  \langle \widetilde r_{L_\tau}\rangle_{\rm ref}
  =
  \frac{
    \operatorname{Cov}_{\rm ref}
    \!\left(\widetilde r_{L_\tau},e^{\rmi\theta}\right)
  }{
    \langle e^{\rmi\theta}\rangle_{\rm ref}
  }\ 
   .
  \label{eq:phase-covariance}
\end{equation}
All quantities on the right-hand side are evaluated using configurations sampled from the reference measure. Hence, the phase-quenched gap ratio provides a quantitatively faithful proxy for its physically reweighted value only when the covariance decreases sufficiently rapidly relative to the average phase, so that the ratio in Eq.~\eqref{eq:phase-covariance} remains small on the scale of the transition-induced spectral response.

Figure~\ref{fig:covariance} displays the interaction dependence of the quantities entering Eq.~\eqref{eq:phase-covariance} for the Haldane-Hubbard model. The average phase decreases rapidly with increasing $U$ and along the accessible $(L,L_\tau)$ sequence [Fig.~\ref{fig:covariance}(b)], reflecting the increasing severity of the phase problem. Nevertheless, the covariance between $\widetilde r_{L_\tau}$ and the determinant phase remains sufficiently small relative to the average phase (typically by an order of magnitude) that the resulting reweighting correction exhibits no resolved systematic growth over the sizes investigated [Figs.~\ref{fig:covariance}(a) and \ref{fig:covariance}(c)]. Within the present statistical resolution, this supports the use of $\langle\widetilde r_{L_\tau}\rangle_{\rm ref}$ as a reliable proxy for locating the transition-sensitive crossover in the Haldane-Hubbard model, even in the regime affected by the phase
problem.

\prlsection{Discussion and Outlook}
Our results establish that local singular-value correlations of the DQMC fermion matrices can retain transition-sensitive information even when observables requiring sign or phase reweighting become difficult to access. Importantly, however, this diagnostic is not independent of the HS representation, since its sensitivity depends on how the auxiliary field encodes the correlations associated with the physical phase. A decoupling that is poorly matched to the relevant ordering channel need not produce a singular-value feature tied to the corresponding critical point, as illustrated by the SU(2)-symmetric decoupling discussed in the SM~\cite{supplementalMaterial}. Here the spin-resolved decomposition explicitly breaks such a symmetry, as does the order parameter in the ordered Mott regime.

This dependence also suggests a systematic route forward. We have generalized Eq.~\eqref{eq:hs-spin-correlation} to channel-matched HS decouplings for which auxiliary-field two-point functions can be related directly to correlations of the targeted physical operator. Such constructions allow the HS representation to be tailored to charge, pairing, bond, or current channels, providing a controlled way to ask whether their ordering phenomena generate corresponding singular-value fingerprints~\cite{chen2026channelmatched}. This turns the freedom in choosing an HS decomposition from a technical choice into a potential design principle for constructing matrix-based diagnostics of many-body order, even in the presence of the ubiquitous sign problem.


\prlsection{Acknowledgments}
Numerical calculations were performed on the Carya and Sabine clusters at the University of Houston's Research Computing Data Core. This work also used ACES at Texas A\&M High Performance Research Computing through allocation PHY240046 from the Advanced Cyberinfrastructure Coordination Ecosystem: Services \& Support (ACCESS) program, which is supported by U.S. National Science Foundation grants 2138259, 2138286, 2138307, 2137603, and 2138296. R.M.~acknowledges that the research was funded in part by the Robert A.~Welch Foundation, Grant \#L-E-0001-19921203.

The data and analysis scripts supporting this work are available in the Texas Data Repository~\cite{repository}.


\bibliographystyle{apsrev4-2}
\bibliography{ref}


\clearpage
\beginsupplement
\onecolumngrid
\vskip 10pt
\begingroup
  \centering
  \phantomsection
  \label{sec:supplemental-material}
  {\large\textbf{SUPPLEMENTAL MATERIAL}}\\[0.4em]
  {\large\textbf{Sign-problem-resilient singular-value probe in determinant quantum Monte Carlo}}\\[0.6em]
  Wen Chen and Rubem Mondaini
\par
\endgroup
\vskip 8.5pt
\twocolumngrid



In these supplementary materials, we expand on the diagnostics used in the main text, provide same-cluster and imaginary-time benchmarks for the singular-value response, characterize the space-time organization of the auxiliary fields, and develop analytically tractable limits that clarify the origin of the low-singular-value behavior. We also examine an alternative SU(2)-invariant Hubbard-Stratonovich decomposition as a negative control for the role played by the decoupling channel.

\section{S1. Determinant quantum Monte Carlo formalism and the reference ensemble}

In finite-temperature determinant quantum Monte Carlo (DQMC), we discretize the imaginary-time direction, $\beta=L_\tau\Delta\tau$, and apply a Trotter decomposition to each time slice, $e^{-\Delta\tau(\hat h_U+\hat h_K)}= e^{-\Delta\tau\hat h_U}\,e^{-\Delta\tau\hat h_K}+{\cal O}(\Delta\tau ^2)$, where $\hat h_U$ and $\hat h_K$ denote the interaction and kinetic parts of the Hamiltonian, respectively. Each interaction factor is then decomposed by the discrete spin-channel Hubbard-Stratonovich (HS) transformation~\cite{hirsch1983discrete},
\begin{equation}
  e^{-\Delta\tau U(\hat n_{i\uparrow}-\tfrac12)(\hat n_{i\downarrow}-\tfrac12)}
  =
  \tfrac12\,e^{-\Delta\tau U/4}
  \sum_{x_{i,\tau}=\pm1}
  e^{\lambda x_{i,\tau}(\hat n_{i\uparrow}-\hat n_{i\downarrow})},
  \label{eq:sm-hs}
\end{equation}
with $\cosh\lambda=e^{\Delta\tau U/2}$ as in the main text, so that $\tanh^2\lambda=1-e^{-\Delta\tau U}$.
After integrating out the fermions, the partition function is rewritten, up to an $x$-independent constant, as $Z_{\Delta\tau}=\sum_x W[x]$, where $W[x]=\prod_\sigma\det M^\sigma[x]$ is the configuration weight of the main text.

To evaluate the sampling weight, the products of single-time-slice propagators entering $M^\sigma[x]$ are stabilized using QR decompositions performed at fixed intervals, together with periodic refresh~\cite{loh2005numerical,bauer2020fast}. Directly extracting the smallest singular values from this long-product representation is nevertheless numerically ill-conditioned because its spectrum spans exponentially separated scales. Following the main text, we therefore use the equivalent space-time fermion matrix $\mathcal M^\sigma[x]$ [Fig.~\ref{fig:models-matrix}(b)]. Its dimension is $N_sL_\tau\times N_sL_\tau$, and the low-end singular spectrum is obtained using distributed PETSc/SLEPc routines~\cite{balay2026petsc,hernandez2005slepc} for sparse matrices, in a process whose computational complexity is much lower than the expected ${\cal O}(N_s^3L_\tau^3)$ for a full singular value decomposition. During the Monte Carlo simulation, we store the auxiliary-field configurations $x\equiv\{x_{i,\tau}\}$. The matrix $\mathcal M^\sigma[x]$ is subsequently reconstructed from each stored configuration, from which we compute its low-lying singular values and the determinant phase. The identity $\det\mathcal M^\sigma[x]=\det M^\sigma[x]$~\cite{blankenbecler1981monte} ensures that the reconstructed determinant phase is the same as that entering the Monte Carlo weight.

For the complex-weight Haldane-Hubbard model, we sample configurations from $W_{\rm ref}[x]=|W[x]|$, as in the main text. For each auxiliary-field configuration $x$, we evaluate the adjacent gap ratios $\widetilde r_n[x]$ of Eq.~\eqref{eq:adjacent-gap-ratio} from the space-time fermion matrix $\mathcal M^\uparrow[x]$ (as in the main text, the spectral analysis is restricted to the spin-up sector and the spin label is suppressed). The ensemble average is taken in two layers: we first average $\widetilde r_n[x]$ over the $N_{\rm low}-2$ gap ratios of the lowest $N_{\rm low}$ singular values within a single configuration, and then average the result over the reference ensemble,
\begin{equation}
  \langle\widetilde r\rangle_{N_{\rm low}}
  =
  \frac{1}{Z_{\rm ref}}\sum_x W_{\rm ref}[x]
  \left(
    \frac{1}{N_{\rm low}-2}
    \sum_{n=1}^{N_{\rm low}-2}\widetilde r_n[x]
  \right),
  \label{eq:sm-ref-ensemble-average}
\end{equation}
with the normalization $Z_{\rm ref}=\sum_x W_{\rm ref}[x]$. The main text reports $\langle\widetilde r\rangle_{L_\tau}$, corresponding to $N_{\rm low}=L_\tau$. $\langle\widetilde r\rangle_{N_{\rm low}}$ is a matrix spectral statistic in the reference ensemble, not a thermodynamic observable obtained from the physical ensemble.

%
\begin{figure}[htbp]
  \centering
  \includegraphics[width=\columnwidth]{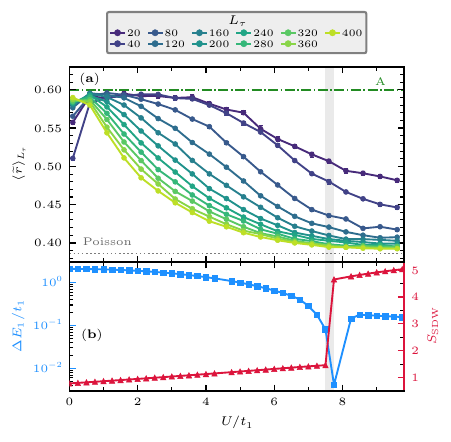}
  \caption{Haldane-Hubbard model on the $L_x=L_y=3$ (18-site) cluster. (a) $\langle\widetilde r\rangle_{L_\tau}$ for the indicated values of $L_\tau$. The dashed-dotted and dotted lines mark the class-A random-matrix benchmark $\langle\widetilde r\rangle_{\rm A}=0.599681$~\cite{kawabata2023singular} and the Poisson value~\cite{atas2013distribution}, respectively. (b) Same-cluster ED results for the many-body gap $\Delta E_1=E_1-E_0$ on the logarithmic left axis and the spin-density-wave structure factor $S_{\rm SDW}$ on the linear right axis. The gray band in both panels marks the ED level-crossing interval $U/t_1\in[7.50,7.75]$. Error bars in panel (a) denote one standard error over independent Markov-chain means.}
  \label{fig:sm-haldane-beta-ed}
\end{figure}

\section{S2. Same-cluster exact-diagonalization reference and imaginary-time extrapolation}
In the main text, we compare the singular-value statistics of the fermion matrices in the reference ensemble with existing estimates of the interaction-driven transition in the Haldane-Hubbard model. These estimates are obtained using methods and geometries that differ from those of the DQMC calculations, including density-matrix renormalization group calculations on finite-width geometries and dynamical mean-field approaches. To provide a benchmark free of this mismatch in finite-size geometry, we additionally perform exact diagonalization (ED) on precisely the same cluster used in our DQMC calculations.

Specifically, we perform ED on the $L_x=L_y=3$ ($N_s=18$) cluster used in Fig.~\ref{fig:models-matrix}(a). Figure~\ref{fig:sm-haldane-beta-ed}(b) shows the excitation gap $\Delta E_1=E_1-E_0$ and the spin-density-wave structure factor,
\begin{equation}
S_{\rm SDW}
=
\frac{1}{N_s}
\sum_{i,j}
(-1)^{\eta}
\left\langle
\left(\hat n_{i,\uparrow}-\hat n_{i,\downarrow}\right)
\left(\hat n_{j,\uparrow}-\hat n_{j,\downarrow}\right)
\right\rangle,
\end{equation}
where $\eta=0$ ($\eta=1$) when sites $i$ and $j$ belong to the same (different) sublattice~\cite{shao2021interplay}. Both quantities change abruptly between $U/t_1=7.50$ and $7.75$, bracketing the finite-cluster level crossing associated with the first-order transition; we therefore use this interval as our same-cluster ED reference. This interval lies above the bulk estimates $U/t_1\in[6.2,6.8]$ quoted in the main text using other numerical methods, reflecting the substantial finite-size shift on the 18-site cluster. As shown in Fig.~\ref{fig:sm-haldane-beta-ed}(a), increasing $L_\tau$ progressively suppresses $\langle\widetilde r\rangle_{L_\tau}$ throughout this same interaction range, motivating the $L_\tau\to\infty$ analysis below.

\begin{figure}[htbp]
  \centering
  \includegraphics[width=\columnwidth]{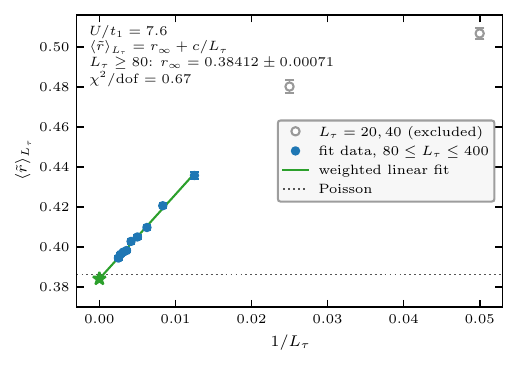}
  \caption{Imaginary-time extrapolation of the low-window gap ratio at the Haldane-Hubbard transition. For $U/t_1=7.6$, inside the same-cluster ED reference interval $[7.50,7.75]$, $\langle\widetilde r\rangle_{L_\tau}$ (with $N_{\rm low}=L_\tau$) is shown against $1/L_\tau$. Gray open circles mark the excluded small grids $L_\tau=20,40$; blue circles are the fit data $80\le L_\tau\le360$; the red diamond marks $L_\tau=400$; the green star marks the weighted linear extrapolation to $L_\tau\to\infty$, $r_\infty=0.3841(7)$ ($\chi^2/\mathrm{dof}=0.67$). The gray dotted line marks the Poisson benchmark $2\ln2-1=0.38629$. Error bars denote one standard error estimated from independent Markov-chain means.}
  \label{fig:sm-transition-extrapolation}
\end{figure}

To determine whether this suppression persists toward the zero-temperature limit, we examine $\langle\widetilde r\rangle_{L_\tau}$ at $U/t_1=7.6$, inside the same-cluster ED reference interval $[7.50,7.75]$, as a function of $1/L_\tau$ in Fig.~\ref{fig:sm-transition-extrapolation}. At fixed $\Delta\tau$, the limit $L_\tau\to\infty$ corresponds to $\beta\to\infty$. Excluding the two smallest grids, $L_\tau=20$ and $40$, an empirical weighted linear fit in $1/L_\tau$ over $80\le L_\tau\le400$ yields $r_\infty=0.38412(71)$, very close to the Poisson benchmark $2\ln2-1=0.38629$ and far below the class-A random-matrix value $\langle\widetilde r\rangle_{\rm A}=0.599681$. The strong suppression near the same-cluster ED transition therefore persists in the $L_\tau\to\infty$ limit and is not a finite-temperature effect.

\begin{figure*}[htbp]
  \centering
  \includegraphics[width=\textwidth]{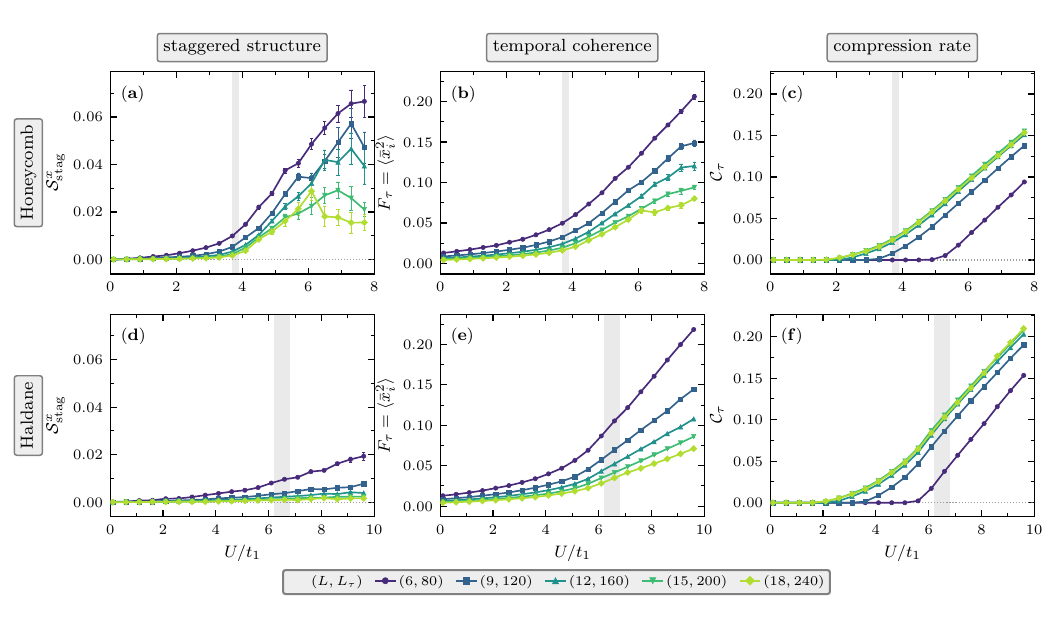}
  \caption{Space-time structure of the DQMC auxiliary field in the honeycomb Hubbard and Haldane-Hubbard models. The top and bottom rows show the two models, respectively. The three columns show the normalized equal-time staggered field correlation $\mathcal S^x_{\rm stag}$ [(a,d)], the zero-frequency temporal coherence $F_\tau$ [(b,e)], and the normalized temporal compressibility $\mathcal C_\tau$ [(c,f)]. Curves are labeled by $(L,L_\tau)$. For the Haldane-Hubbard model, all quantities are evaluated in the reference ensemble $W_{\rm ref}=|W|$, without phase reweighting. Gray bands mark the reference transition intervals $U/t_1\in[3.7,3.9]$ and $[6.2,6.8]$ for the honeycomb Hubbard and Haldane-Hubbard models, respectively. Error bars denote one standard error estimated from independent Markov-chain means.}
  \label{fig:sm-field-structure}
\end{figure*}

\section{S3. Auxiliary-field structure and interaction dependence}

Since the space-time fermion matrix $\mathcal M^\sigma[x]$ is constructed directly from the Hubbard-Stratonovich (HS) field configuration $x=\{x_{i,\tau}\}$, changes in the spatial and imaginary-time organization of this field can therefore reorganize the singular-value spectrum of $\mathcal M^\sigma$. The purpose of this section is to characterize this underlying field structure as the interaction strength is increased. We will define quantities that probe complementary aspects of the organization of the HS configurations from which the fermion matrices are constructed.

We first quantify spatial antiferromagnetic correlations within each imaginary-time slice through
\begin{equation}
  \mathcal S^x_{\rm stag}[x]
  =
  \frac{1}{L_\tau N_s(N_s-1)}
  \sum_{\tau=1}^{L_\tau}
  \sum_{i\neq j}
  \eta_i\eta_j x_{i,\tau}x_{j,\tau}\ ,
  \label{eq:sm-field-staggered-pairs}
\end{equation}
where $\eta_i=\pm1$ denotes the sublattice sign of site $i$. Thus, $\mathcal S^x_{\rm stag}$ is the equal-time staggered correlation averaged over all distinct pairs of spatial sites. It vanishes on average for spatially independent, unbiased fields and reaches unity for a perfectly staggered configuration, $x_{i,\tau}=\eta_i$. This quantity therefore isolates the spatial staggering of the auxiliary field.

Spatial organization does not exhaust the information entering $\mathcal M^\sigma[x]$, since the ordering of the HS variables along imaginary time is also retained by the space-time matrix. We characterize this temporal structure by
\begin{equation}
  F_\tau[x]
  =
  \frac{1}{N_sL_\tau^2}
  \sum_{i=1}^{N_s}
  \sum_{\tau,\tau'=1}^{L_\tau}
  x_{i,\tau}x_{i,\tau'}\ .
  \label{eq:sm-field-temporal-correlation}
\end{equation}
Hence, $F_\tau$ measures the time-integrated autocorrelation of the HS field at a fixed spatial site, or equivalently its zero-Matsubara-frequency component. For an independent Ising field, only the $\tau=\tau'$ terms survive on average and
$\langle F_\tau\rangle_{\rm iid}=1/L_\tau$. A field that is completely uniform along imaginary time instead has $F_\tau=1$. Increasing $F_\tau$ therefore signals increasing persistence of the HS field along the imaginary-time direction.

The quantity $F_\tau$ probes only the zero-frequency component and can remain small for temporally organized sequences whose positive and negative portions cancel in the time average. We therefore complement it with the normalized temporal compressibility
\begin{equation}
  \mathcal C_\tau[x]
  =
  \frac{\ell_{\rm iid}-\ell[x]}
       {\ell_{\rm iid}-\ell_{\rm rep}},
  \label{eq:sm-field-compression}
\end{equation}
where $\ell[x]$ is the lossless compressed length of the binary representation of configuration $x$. Before compression, the field is ordered site by site so that each site's imaginary-time history is contiguous and encoded with one bit per Ising variable. The resulting bitstream is compressed using raw DEFLATE, the lossless compression scheme underlying gzip~\cite{deutsch1996deflate}. The reference $\ell_{\rm iid}$ is the mean compressed length of size-matched fields drawn independently at every space-time point, whereas $\ell_{\rm rep}$ is obtained from random spatial configurations repeated identically along imaginary time. This normalization gives $\langle\mathcal C_\tau\rangle_{\rm iid}=0$ and $\langle\mathcal C_\tau\rangle_{\rm rep}=1$. Thus, unlike $F_\tau$, $\mathcal C_\tau$ is also sensitive to temporal redundancy such as repeated motifs or extended domains even when the time average of the field is small.

Figure~\ref{fig:sm-field-structure} shows that the HS configurations become progressively more organized with increasing interaction strength in both models. For the honeycomb Hubbard model, the sampled ensemble is the physical one, and the growth of $\mathcal S^x_{\rm stag}$ across the AFM Mott transition directly reflects the development of staggered spin correlations through the HS-spin relation of Eq.~\eqref{eq:hs-spin-correlation} in the main text. The concomitant increase of $F_\tau$ and $\mathcal C_\tau$ shows that this spatial organization is accompanied by increasing coherence along imaginary time. The Haldane-Hubbard model also develops AFM Mott order and is decoupled in the same symmetry-compatible spin channel. Here, however, the field diagnostics are evaluated in the phase-quenched reference ensemble, $W_{\rm ref}=|W|$, so their correspondence with physical spin correlations requires phase reweighting. Their comparatively smooth evolution across the physical transition therefore does not signal a mismatch of the HS channel. Moreover, the same-sublattice Haldane hopping competes with the simple bipartite antiferromagnetic tendency, reducing the degree of staggered field organization, although imaginary-time organization is comparable to, if not more pronounced than, in the pure $t_2=0$ case. 

This distinction is central to interpreting the singular-value results. Matching the symmetry of the HS decomposition to that of the emerging order allows the associated field organization to be encoded in the space-time fermion matrix, but it does not imply that every field observable measured in the reference ensemble must itself locate the physical transition, since phase-reweighting corrections are observable-dependent. In particular, Fig.~4 of the main text shows that the correction is small for the adjacent singular-value gap ratio, so that $\langle\widetilde r\rangle_{\rm ref}$ retains the transition-sensitive behavior even though the reference-ensemble field diagnostics in Fig.~\ref{fig:sm-field-structure} do not sharply identify the transition.


\section{S4. Symmetry classes and mechanisms of the low singular-value organization}

In this section, we first identify the symmetry classes that govern the singular-value statistics of the space-time fermion matrix $\mathcal M^\sigma[x]$ and the corresponding random-matrix benchmarks used in the main text. We then examine two analytically tractable limits of the HS field that help explain why the low end of the singular-value spectrum is especially sensitive to the developing field structure. A static staggered field connects the lowest singular values to the smallest single-particle scales, whereas a field that is uniform along imaginary time produces degenerate or nearly degenerate pairs between the $\pm q$ sectors (defined below), leading to a suppression of the adjacent gap ratio. The constructions below are exact within the limits to which they are derived and provide a simple interpretation of the numerical trends observed in the main text.

We start by noticing that the symmetry class of $\mathcal M^\sigma[x]$ follows directly from the kinetic matrix $K$. For the honeycomb Hubbard model, $K$ is real, and since the HS `potential' $V_\tau=\mathrm{diag}(x_{i,\tau})$ we use is also real~\cite{hirsch1983discrete}, each slice propagator $B_\tau^\sigma=e^{\sigma\lambda V_\tau}e^{-\Delta\tau K}$ and the resulting space-time matrix $\mathcal M^\sigma[x]$ are real. The latter therefore belongs to the non-Hermitian class AI. In the Haldane-Hubbard model, the complex next-nearest-neighbor hopping makes $K$, and hence $\mathcal M^\sigma[x]$, generically complex, placing it in class A. The singular values can equivalently be viewed as the positive eigenvalues of the Hermitized matrix
\begin{equation}
  \mathscr D^\sigma[x]
  =
  \begin{pmatrix}
    0 & \mathcal M^\sigma[x]\\
    \mathcal M^{\sigma\dagger}[x] & 0
  \end{pmatrix},
\end{equation}
whose spectrum is $\{\pm s_j\}$. Under this Hermitization, classes AI and A map to BDI and AIII, respectively~\cite{kawabata2023singular}, yielding the random-matrix benchmarks $\langle\widetilde r\rangle_{\rm BDI}=0.530654$ and $\langle\widetilde r\rangle_{\rm AIII}=0.599681$ used in the main text.

\subsection{Origin of the low singular-value scale}
We are now in a position to address the first mechanism: why is the low end of the singular-value spectrum particularly sensitive to the developing field structure? For that, consider a perfectly staggered HS field that is uniform along imaginary time, $V_\tau=\Gamma=\mathrm{diag}(\eta_i)$ ($\eta_i=\pm1$ sublattice signs), with $\Gamma^2=I$, for a bipartite nearest-neighbor model satisfying $\{K,\Gamma\}=0$. In this case, an eigenstate $|a\rangle$ of $K$ with eigenvalue $\xi_a>0$ is paired with $\Gamma|a\rangle$ at $-\xi_a$, since $K\Gamma|a\rangle=-\Gamma K|a\rangle=-\xi_a\Gamma|a\rangle$. These two states span a subspace invariant under both $K$ and $\Gamma$, in which one may write $K_a=\xi_a\sigma_z$ and $\Gamma_a=\sigma_x$. The single-slice propagator restricted to this subspace is therefore $B_a=e^{\sigma\lambda\sigma_x}e^{-\Delta\tau\xi_a\sigma_z}$.

Both exponential factors have unit determinant, so $\det B_a=1$, while $\mathrm{Tr}\,B_a=2\cosh\lambda\cosh(\Delta\tau\xi_a)$. Thus, if $b_a^+$ and $b_a^-$ are the two eigenvalues, they obey $b_a^+b_a^-=1$ and $b_a^++b_a^-=2\cosh\lambda\cosh(\Delta\tau\xi_a)$. They can consequently be parameterized as
\begin{equation}
  b_a^\pm=e^{\pm\gamma_a},
  \qquad
  \cosh\gamma_a
  =
  \cosh\lambda\,\cosh(\Delta\tau\xi_a).
  \label{eq:sm-gamma}
\end{equation}
For fixed $\lambda$, $\gamma_a$ increases monotonically with $\xi_a\geq0$. Thus, the single-particle states closest to the band center correspond to the smallest values of $\gamma_a$. We next show how $\gamma_a$ enters the low singular values of the space-time matrix.

The uniformity of the field along imaginary time also makes the space-time matrix translation invariant in the temporal direction [see Fig.~1(b) in the main text]. To make this explicit, consider a vector whose component on time slice $\tau$ has the form $\Psi_\tau=e^{\rmi q\tau}\boldsymbol u$, where $\boldsymbol u$ is an $N_s$-dimensional spatial vector. Since the action of the space-time matrix away from the temporal boundary is $\Psi_\tau-B\Psi_{\tau-1}$, such a mode is acted on by the
$N_s\times N_s$ block 
\begin{equation}
  \mathcal M(q)=I-e^{-\rmi q}B.
  \label{eq:sm-Mq}
\end{equation}
The antiperiodic boundary condition connecting the first and last time slices requires $e^{\rmi qL_\tau}=-1$, and hence $q_n=(2n+1)\pi/L_\tau$. The temporal Fourier transformation therefore decomposes the full $N_sL_\tau$-dimensional matrix into $L_\tau$ independent blocks $\mathcal M(q_n)$. Since this transformation is unitary, the singular values of the time-uniform space-time matrix are the union of the singular values of these $q$ sectors.

A closed form for these singular values is obtained in the symmetric gauge. The slice propagator is then $\widetilde B=e^{-\Delta\tau K/2}e^{\sigma\lambda\Gamma} e^{-\Delta\tau K/2}$, which is Hermitian positive definite and is similar to $B$, with the same eigenvalues $e^{\pm\gamma_a}$. Correspondingly, the $q$ block becomes $\widetilde{\mathcal M}(q)=I-e^{-\rmi q}\widetilde B$. In the orthonormal eigenbasis of $\widetilde B$, this matrix is diagonal, with entries  $1-e^{\pm\gamma_a}e^{-\rmi q}$. Its singular values are therefore the moduli of these entries,
\begin{equation}
  \widetilde s_{a,q}^{<}
  =
  \left|1-e^{-\gamma_a}e^{-\rmi q}\right|,
  \qquad
  \widetilde s_{a,q}^{>}
  =
  e^{\gamma_a}\widetilde s_{a,q}^{<}.
  \label{eq:sm-branches}
\end{equation}

At the low end, where both $\gamma_a$ and $q$ are small,
\begin{equation}
  \left(\widetilde s_{a,q}^{<}\right)^2
  =
  \left(1-e^{-\gamma_a}\right)^2
  +2e^{-\gamma_a}(1-\cos q)
  \simeq
  \gamma_a^2+q^2 .
  \label{eq:sm-low-branch}
\end{equation}
Equation~\eqref{eq:sm-low-branch} makes the origin of the low-energy sensitivity explicit: the lowest branch is controlled jointly by the smallest $\gamma_a$, inherited from the single-particle states closest to the band center, and the smallest antiperiodic momentum, $q_{\min}=\pi/L_\tau$. Changes in the spatial and imaginary-time organization of the HS field therefore enter most directly at the hard
edge of the singular-value spectrum.

A few qualifications are important. Equations~\eqref{eq:sm-branches} and~\eqref{eq:sm-low-branch} are exact in the symmetric gauge for the perfectly staggered, time-uniform bipartite limit. The transformation from the slice propagator used in the simulations to the symmetric gauge is a nonunitary similarity transformation, which preserves the eigenvalues of $B$ but not, in general, the singular values of the full space-time matrix. We therefore use this limit to expose the scales governing the hard edge rather than as an exact expression for the numerical spectrum. Exact zero modes $\xi_a=0$ can be treated by first diagonalizing $\Gamma$ within the zero-mode subspace. Finally, the same-sublattice Haldane hopping breaks $\{K,\Gamma\}=0$, so the argument is exact for the nearest-neighbor bipartite limit and serves only as a reference mechanism for the Haldane-Hubbard case.

This first mechanism identifies why the low singular values are especially sensitive to the developing field structure, but it does not, by itself, explain why their adjacent gap ratio is suppressed. We turn next to the temporal near-doublet mechanism responsible for that effect.

\subsection{Temporal near-doublets and gap-ratio suppression}

We now consider the second mechanism responsible for the suppression of the adjacent gap ratio. Let $\mathcal M_0$ denote the space-time fermion matrix in the idealized time-uniform limit $B_\tau=B$ for all $\tau$. In this limit, $\mathcal M_0$ is translationally invariant along imaginary time and decomposes into the antiperiodic sectors $q_n=(2n+1)\pi/L_\tau$ introduced above. In the symmetric gauge, Eq.~\eqref{eq:sm-branches} shows that the singular values depend on the temporal momentum only through $\cos q$, and therefore
\begin{equation}
  \widetilde s_{a,-q}^{\lessgtr}
  =
  \widetilde s_{a,q}^{\lessgtr}\ .
  \label{eq:sm-doublet-pairing}
\end{equation}
The antiperiodic sectors $q$ and $-q$ thus form exact singular-value doublets in this limit. For a real slice propagator, as in the honeycomb Hubbard model, the same pairing holds directly in the original gauge because the two blocks satisfy $\mathcal M_0(-q)=\mathcal M_0(q)^*$ and therefore have identical singular values. For a generic complex propagator, as in the Haldane-Hubbard model, the nonunitary transformation to the symmetric gauge does not preserve singular values, so Eq.~\eqref{eq:sm-doublet-pairing} should instead be regarded as an analytically tractable reference limit.

The connection to the gap ratio becomes transparent once the doublets are weakly split. Let the $j$th pair have center $c_j$ and splitting $\delta_j$, so that its two singular values are $c_j\pm\delta_j/2$, and let $D_j=c_{j+1}-c_j$ denote the separation between neighboring pair centers. The consecutive spacings then alternate between the intra-doublet splitting $\delta_j$ and an inter-doublet spacing of order $D_j$. When $\delta_j\ll D_j$, the corresponding adjacent gap ratios scale schematically as
\begin{equation}
  \widetilde r \sim \frac{\delta_j}{D_j}\ll1.
  \label{eq:sm-doublet-gap-ratio}
\end{equation}
Thus, an approximate pairing of neighboring singular values naturally suppresses $\widetilde r$ by producing an alternation of small and large gaps. Indeed, a direct example of this regime is the values of the ratio of adjacent gaps in the synthetic ensemble [Fig.~2 of the main text] when the parametrization $\alpha=1$. In this case, the perfectly staggered field in real space that is propagated along imaginary time makes the singular-value-doublets degenerate ($\delta_j=0$), thereby preventing the observation of proper level repulsion.

Instead, for an actual fluctuating HS configuration, $B_\tau$ depends on $\tau$ and imaginary-time translation invariance is lost. The temporal Fourier sectors are then coupled, lifting the $q$--$(-q)$ degeneracy and producing a finite splitting $\delta_j$. As the HS field becomes more coherent along imaginary time, the departures from the time-uniform limit are reduced, and the near-doublet structure becomes increasingly relevant. This provides a direct connection with the growth of the temporal diagnostics $F_\tau$ and $\mathcal C_\tau$ discussed in Sec.~S3: increasing temporal organization reduces the mixing between time sectors and can suppress the local singular-value repulsion measured by $\langle\widetilde r\rangle$.

\begin{figure*}[htbp]
  \centering
  \includegraphics[width=0.9\textwidth]{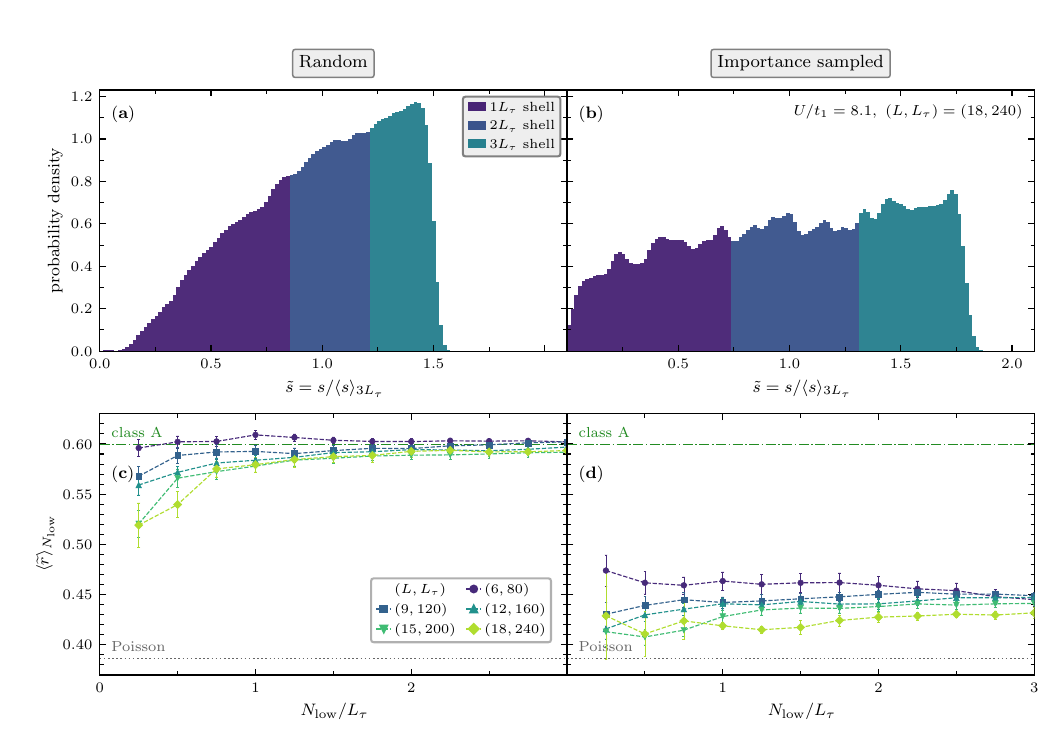}
  \caption{Haldane-Hubbard successive $L_\tau$-sized rank blocks and cumulative low-window statistics at $U/t_1=8.1$. Top row: the complete lowest $3L_\tau$ sector for $(L,L_\tau)=(18,240)$, normalized in each realization by the mean of those $3L_\tau$ values, for (a) independent equiprobable fields and (b) importance-sampled fields. Each fine histogram bin gives the total probability density; its color identifies the $1L_\tau$, $2L_\tau$, or $3L_\tau$ rank block with the largest local contribution. The upper panels use five independent randomized realizations and three independent Markov-chain means, respectively. Bottom row: mean adjacent-gap ratio computed from the lowest $N_{\rm low}$ singular values as $N_{\rm low}/L_\tau$ is increased from $0.25$ to $3$, for (c) random fields and (d) importance-sampled fields; each curve corresponds to one $(L,L_\tau)$ size. The horizontal dash-dotted and dotted lines mark the class-A random-matrix benchmark and the Poisson value, respectively. Error bars denote the standard error over independent realizations or Markov-chain means, respectively.}
  \label{fig:sm-random-vs-importance}
\end{figure*}

\section{S5. Window dependence of the low singular-value statistics}


\begin{figure*}[htbp]
  \centering
  \includegraphics[width=\textwidth]{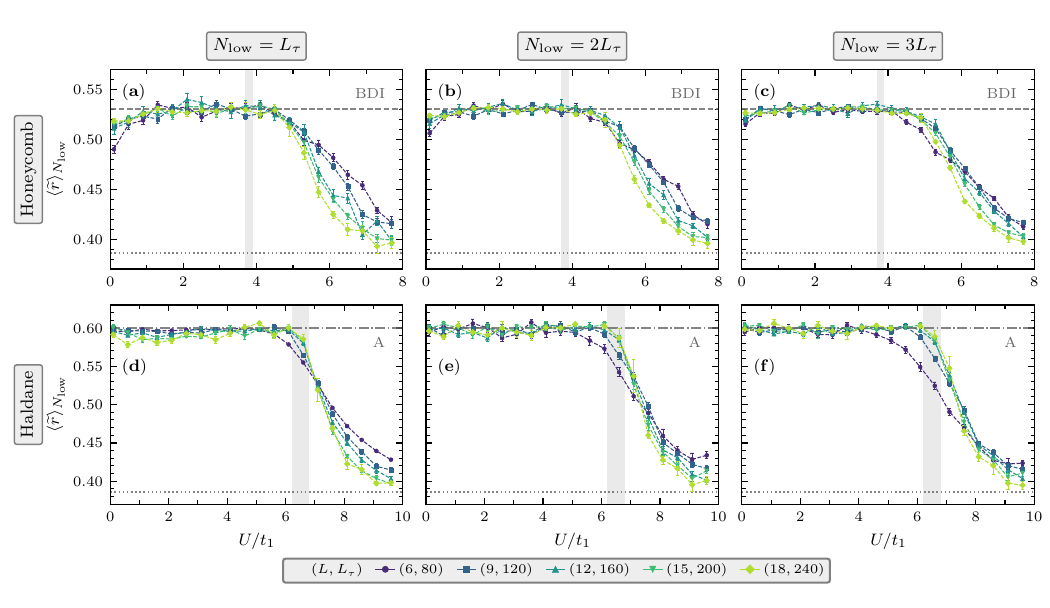}
  \caption{Window-size dependence of the adjacent gap ratio in the low singular-value sector. The top and bottom rows show the honeycomb Hubbard and Haldane-Hubbard models, respectively, over $0\le U/t_1\le8$ and $0\le U/t_1\le10$. The three columns use the lowest $N_{\rm low}=L_\tau$ [(a,d)], $2L_\tau$ [(b,e)], and $3L_\tau$ [(c,f)] singular values to compute $\langle\widetilde r\rangle_{N_{\rm low}}$. Curves are labeled by $(L,L_\tau)$. The Haldane-Hubbard results are evaluated in the reference ensemble $W_{\rm ref}=|W|$ without phase reweighting. Gray bands mark the reference intervals $U/t_1\in[3.7,3.9]$ and $[6.2,6.8]$, respectively. Horizontal lines mark the Poisson benchmark and the BDI and class-A random-matrix benchmarks for the top and bottom rows, respectively. Error bars denote one standard error estimated from independent Markov-chain means.}
  \label{fig:sm-window-audit}
\end{figure*}

In the time-uniform limit of the mechanism above, each spatial mode $a$ produces $L_\tau$ singular values, one per imaginary-time momentum sector. The low-end window $N_{\rm low}=L_\tau$ is therefore of the same order as the temporal-multiplet scale: a conservative choice inspired by this structure, since for nonuniform time slices the low-end windows of different spatial modes can interlace, and the multiplet structure of the time-uniform limit is only approximate. 

Figure~\ref{fig:sm-random-vs-importance} examines this question from the spectral side in the Haldane model at fixed $U/t_1=8.1$. Here, the randomized-field spectrum is highly nonuniform at its low end, so that the smallest window $N_{\rm low}=L_\tau$ sits on this hard edge and deviates from the statistical benchmark. As the window is enlarged toward $3L_\tau$, the cumulative averages relax back toward the benchmark (bottom row). The residual downward drift of the randomized-field curve in Fig.~\ref{fig:benchmark}(c) of the main text therefore reflects this hard-edge nonuniformity and its accumulation over the window. 

The importance-sampled suppression, by contrast, does not relax back to the baseline as the window grows because it originates from field reorganization. To confirm that the window choice does not drive the conclusions, Fig.~\ref{fig:sm-window-audit} compares $\langle\widetilde r\rangle_{N_{\rm low}}(U)$ for $N_{\rm low}=L_\tau$, $2L_\tau$, and $3L_\tau$. The onset of the departure from the respective baselines (BDI for honeycomb, A for Haldane) is essentially independent of the window size, showing no dramatic change across the three windows. The choice $N_{\rm low}=L_\tau$ is therefore robust and conservative, and the singular value conclusions do not depend on the window definition.

\section{S6. SU(2)-invariant Hubbard-Stratonovich decomposition as a negative control}

As a negative control, we repeat the honeycomb benchmark with a different HS decomposition for the same physical transition. That is, on the half-filled honeycomb Hubbard model, we replace the Ising spin-channel HS transformation~\cite{hirsch1983discrete} with the SU(2)-invariant auxiliary-field decomposition~\cite{assaad1999su2} and probe the same AFM Mott transition. If the singular value response appears only in the spin-channel decomposition, it indicates that it is tied to the auxiliary-field structure produced by that particular decomposition rather than to the transition itself.

Under the SU(2)-invariant decomposition, the auxiliary field takes four values, $x_{i,\tau}\in\{-2,-1,1,2\}$, and no longer encodes the Ising spin channel. Following the four-field quadrature introduced in
Ref.~\cite{assaad1999su2}, we approximate the local interaction factor, for a finite imaginary-time step, as
\begin{equation}
\begin{aligned}
  e^{-\Delta\tau U(n_{i,\tau}-1)^2/2}
  &=
  \frac{1}{4}
  \sum_{x=\pm1,\pm2}
  \gamma(x)\,
  e^{\lambda\eta(x)(n_{i,\tau}-1)}
  \\
  &\quad
  +\mathcal O\!\left[(\Delta\tau U)^4\right]\ ,
\end{aligned}
\label{eq:sm-su2-decomposition}
\end{equation}
with $n_{i,\tau}=n_{i,\tau,\uparrow}+n_{i,\tau,\downarrow}$ the site occupation, $\lambda=\rmi\sqrt{\Delta\tau U/2}$, $\eta(\pm1)=\pm\sqrt{2(3-\sqrt{6})}$, $\eta(\pm2)=\pm\sqrt{2(3+\sqrt{6})}$, and $\gamma(\pm1)=1+\sqrt{6}/3$, $\gamma(\pm2)=1-\sqrt{6}/3$. For a fixed auxiliary-field configuration, the factors $\gamma(x_{i,\tau})/4$ contribute to the auxiliary-field weight, while the
fermions couple to the field through the complex density-channel factor $e^{\lambda\eta(x_{i,\tau})(n_{i,\tau}-1)}$. Since $\lambda=\rmi\sqrt{\Delta\tau U/2}$ for $U>0$, the corresponding one-body propagators are complex even for the honeycomb kinetic matrix, so the SU(2) spectrum is benchmarked against class A.

\begin{figure}[htbp]
  \centering
  \includegraphics[width=\columnwidth]{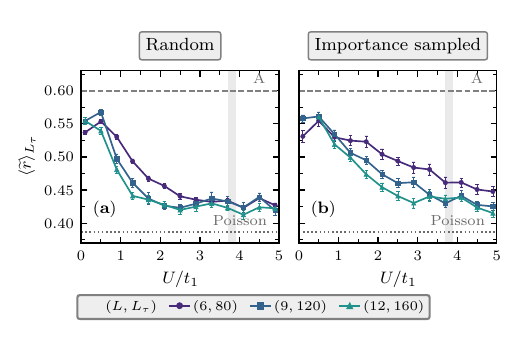}
  \caption{Adjacent gap ratio in the low singular-value sector of the half-filled honeycomb Hubbard model with the SU(2)-invariant HS transformation. (a) Randomized auxiliary fields. (b) Importance-sampled auxiliary fields. The gray band marks the reference interval $U/t_1\in[3.7,3.9]$, and the horizontal dashed (dotted) line marks the class-A random-matrix (Poisson) benchmark. Error bars denote one standard error, estimated from independent realizations in (a) and from independent Markov-chain means in (b).}
  \label{fig:sm-su2-gapratio}
\end{figure}

Figure~\ref{fig:sm-su2-gapratio} shows the interaction-strength dependence of the low-lying singular-value statistics, using the lowest $L_\tau$ singular values. In contrast to the spin-channel decomposition, the randomized-field ensemble does not approach the corresponding random-matrix benchmark. For repulsive interactions, the SU(2)-invariant HS field enters each time-slice propagator through a diagonal unitary factor, $D_\tau=\operatorname{diag}[e^{\lambda\eta(x_{i,\tau})}]$, so that $B_\tau=D_\tau e^{-\Delta\tau K}$. Thus, even when the auxiliary fields are independently randomized, the resulting space-time matrices form a highly constrained random-phase ensemble rather than a generic class-A random-matrix ensemble. The absence of agreement with the random-matrix benchmark is therefore not unexpected and further illustrates that the symmetry class alone does not determine the singular-value statistics of the structured DQMC matrices. Indeed, in Fig.\ref{fig:sm-su2-gapratio}(b), the importance-sampled fields also do not track the $U_c/t\in[3.7,3.9]$ transition. 

\begin{figure*}[htbp!]
  \centering
  \includegraphics[width=\textwidth]{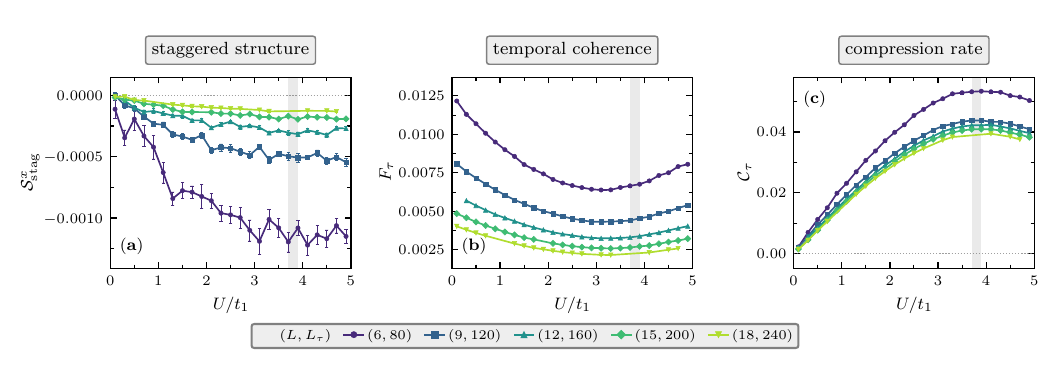}
  \caption{Space-time structure of the SU(2)-invariant HS auxiliary field in the half-filled honeycomb Hubbard model. (a) Normalized staggered field structure $\mathcal S^x_{\rm stag}$. (b) Imaginary-time coherence $F_\tau$. (c) Normalized temporal compression rate $\mathcal C_\tau$. The SU(2) auxiliary field takes values $x_{i,\tau}\in\{-2,-1,1,2\}$. The spatial and temporal metrics are normalized by the field norm of each configuration, while the compression rate uses size-matched independent and imaginary-time-repeated SU(2) fields as zero- and unit-reference. Curves compare $(L,L_\tau)=(6,80)$, $(9,120)$, $(12,160)$, $(15,200)$, and $(18,240)$. The gray band marks the reference interval $U/t_1\in[3.7,3.9]$. Error bars denote one standard error estimated from independent Markov-chain means.}
  \label{fig:sm-su2-field-structure}
\end{figure*}

To understand this, let us focus on the field-structure metrics [Eqs.~\eqref{eq:sm-field-staggered-pairs}-\eqref{eq:sm-field-compression}], now applied to the four-valued field of the SU(2) HS transformation. Figure \ref{fig:sm-su2-field-structure} shows behavior that contrasts sharply with the spin channel: the staggered structure factor does not develop a positive signal over the entire range of $U$ (in the spin channel it grows and accelerates across $[3.7,3.9]$). This is consistent with the identity of Eq.~\eqref{eq:hs-spin-correlation} of the main text, which ties staggered field correlations to the physical staggered spin correlations in the spin-channel decomposition. The SU(2) field does not encode the Ising spin channel, so no such relation holds. The SU(2)-invariant auxiliary field is directly related to charge, rather than spin, correlations. At half filling, the interaction-driven transition is into an antiferromagnetic Mott insulator without accompanying charge-density-wave order. The corresponding HS-field correlations may therefore evolve with interaction strength through the suppression of charge fluctuations, but they are not expected to acquire the staggered critical structure of the AFM order parameter. With that, the SU(2) field therefore does not develop the staggered, $\tau$-uniform pattern of the spin channel, and the two-branch mechanism of Sec.~S4 does not apply to it.

Together with Fig.~\ref{fig:sm-su2-field-structure}, this indicates that the decrease of $\langle\widetilde r\rangle$ in the spin channel corresponds to its characteristic staggered, $\tau$-uniform field organization, which the SU(2) field lacks. This negative evidence establishes only the decomposition specificity of the spin-channel response: it does not exclude that the SU(2) or charge channels exhibit their own singular-value responses at their respective transitions, which are outside the scope of this letter~\cite{chen2026channelmatched}.

\section{S7. Higher-order correlations of the low-lying singular values}

\begin{figure*}[b!]
  \centering
  \includegraphics[width=\textwidth]{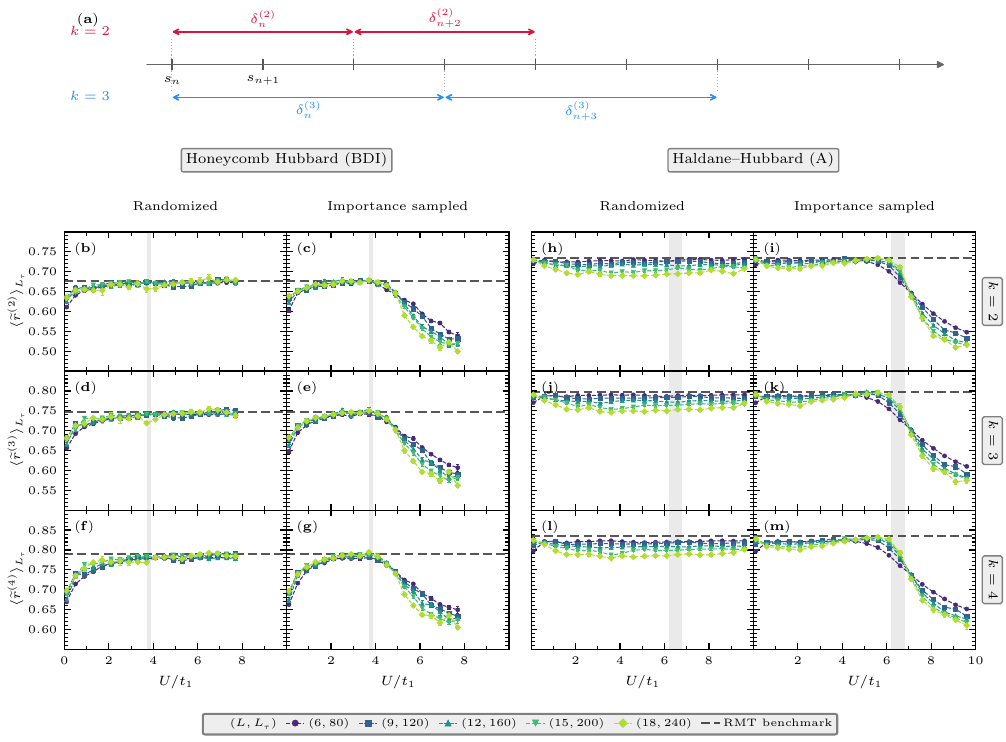}
  \caption{Higher-order correlations of the low-lying singular values. (a) Construction of the nonoverlapping order-$k$ gaps entering $\widetilde r_n^{(k)}$, illustrated for $k=2$ (red) and $k=3$ (blue); the black ticks denote schematically ordered singular values. Panels (b--g) show the half-filled honeycomb Hubbard model, while panels (h--m) show the half-filled Haldane-Hubbard model. Within each model block, rows correspond to $k=2$, $3$, and $4$ from top to bottom, and the left and right columns show independently randomized and importance-sampled auxiliary fields, respectively. All data panels use the lowest $N_{\rm low}=L_\tau$ singular values, with curves labeled by $(L,L_\tau)$. The Haldane--Hubbard importance-sampled results are evaluated in the phase-quenched reference ensemble $W_{\rm ref}=|W|$, without phase reweighting. Gray bands mark the reported transition intervals $U_c/t_1\in[3.7,3.9]$ and $[6.2,6.8]$ for the honeycomb and Haldane-Hubbard models, respectively. Horizontal dashed lines denote the corresponding higher-order random-matrix benchmarks. From top to bottom ($k=2,3,4$), these are $0.6762$, $0.7464$, and $0.7902$ for BDI, and $0.7335$, $0.7964$, and $0.8344$ for class A, as obtained from the scaling relation of Ref.~\cite{tekur2018higher}. Error bars denote one standard error over independent randomized-field realizations or independent Markov-chain means, respectively.}
  \label{fig:sm-higher-order}
\end{figure*}

The main text uses the adjacent gap ratio, formed from three consecutive low-lying singular values, to characterize the shortest-range spectral correlations~\cite{Oganesyan2007,atas2013distribution}. We ask whether the observed interaction response is confined to this local structure or instead reflects a broader reorganization of the low-lying spectrum involving longer-range correlations. To address this question, we turn to higher-order gap ratios, which provide an intuitive probe and, like the adjacent ratio, require no spectral unfolding~\cite{tekur2018higher}. Specifically, for ordered singular values $s_1\le s_2\le\ldots$, we define
\begin{equation}
  \widetilde r_n^{(k)}
  =
  \frac{
    \min\!\left(\delta_n^{(k)},\delta_{n+k}^{(k)}\right)
  }{
    \max\!\left(\delta_n^{(k)},\delta_{n+k}^{(k)}\right)
  }\ .
  \label{eq:sm-higher-order-ratio}
\end{equation}
Here the order-$k$ gap is $\delta_n^{(k)}=s_{n+k}-s_n$. As illustrated in Fig.~\ref{fig:sm-higher-order}(a), the two gaps entering Eq.~\eqref{eq:sm-higher-order-ratio} each span $k$ consecutive nearest-neighbor spacings, with no spacing shared between them. By construction, Eq.~\eqref{eq:sm-higher-order-ratio} recovers the adjacent gap ratio at $k=1$. To extend the comparison beyond this nearest-neighbor case, we examine $k=2,3,4$ using exactly the same auxiliary-field configurations, low-lying spectra, and averaging procedure as in the main text. In particular, for a Wigner-Dyson ensemble with Dyson index $\beta$~\cite{atas2013distribution}, the nonoverlapping order-$k$ ratio follows the effective index
\begin{equation}
  \beta_k=\frac{k(k+1)}{2}\beta+(k-1)
\end{equation}
of Ref.~\cite{tekur2018higher}, such that $P^k(r,\beta)=P(r,\beta_k)$. Consequently, the corresponding mean symmetrized-ratio benchmark increases with $k$.

Following the comparison in Fig.~\ref{fig:benchmark}, we begin with the sign-problem-free honeycomb Hubbard model. Figure~\ref{fig:sm-higher-order}(b--g) shows $k=2,3,4$ from top to bottom, with the randomized-field and importance-sampled results in the left and right columns, respectively. Across all three rows, the two ensembles develop distinct interaction trends: the randomized-field curves rise initially and then settle to nearly constant values, whereas the importance-sampled curves turn downward around and beyond the established AFM transition interval $U/t_1\in[3.7,3.9]$~\cite{sorella2012absence,assaad2013pinning,parisentoldin2015fermionic}. Compared with the adjacent-gap result in Fig.~\ref{fig:benchmark}(b), the higher-order ratios develop a clearer and more coherent downward trend beyond this interval, particularly for the larger systems.

We then turn to the Haldane-Hubbard model, where the auxiliary-field weights are generally complex [Fig.~\ref{fig:sm-higher-order}(h--m)]. The importance-sampled results are evaluated in the phase-quenched reference ensemble $W_{\rm ref}=|W|$, without phase reweighting, and therefore characterize higher-order spectral correlations of the reference matrix ensemble rather than observables of the physical complex-weight ensemble. Here, despite some size dependence, the randomized-field curves remain comparatively smooth functions of $U$, whereas the reference-ensemble curves develop a pronounced downward crossover near the reported transition interval $U/t_1\in[6.2,6.8]$~\cite{vanhala2016topological,he2024phase}.

Taken together, Figs.~\ref{fig:sm-higher-order}(b--g) and \ref{fig:sm-higher-order}(h--m) show that the transition-sensitive response persists at higher orders in both models. To understand what this persistence implies for the spectral reorganization, we view it in light of the near-doublets discussed in Sec.~S4. Whereas the adjacent ratio responds directly to the alternation of small and large spacings produced by near-doublets, a higher-order gap combines several consecutive spacings and is therefore less sensitive to the splitting of any single pair. The continued visibility of the response in the higher-order ratios therefore indicates that the associated spectral correlations extend across a broader interval rather than being confined to individual near-doublets.

\end{document}